**Safety Perspective on Assisted Lane Changes: Insights from Open-Road, Live-Traffic Experiments**

**Konstantinos Mattas***
European Commission, Joint Research Centre, Ispra, VA, Italy
Konstantinos.MATTAS@ec.europa.eu

**Sandor Vass**
European Commission, Joint Research Centre, Ispra, VA, Italy
Sandor.VASS@ec.europa.eu

**Gergely Zachár**
Institute for Software Integrated Systems, Vanderbilt University, Nashville, TN, United States
gergely.zachar@Vanderbilt.Edu

**Junyi Ji**
Institute for Software Integrated Systems, Vanderbilt University, Nashville, TN, United States
junyi.ji@Vanderbilt.Edu

**Derek Gloudemans**
Institute for Software Integrated Systems, Vanderbilt University, Nashville, TN, United States
derek.gloudemans@vanderbilt.edu

**Davide Maggi**
European Commission, Joint Research Centre, Ispra, VA, Italy
Davide.MAGGI@ec.europa.eu

**Akos Kriston**
European Commission, Joint Research Centre, Ispra, VA, Italy
Akos.KRISTON@ec.europa.eu

**Mohamed Brahmi**
European Commission, Directorate General for Internal Market, Industry, Entrepreneurship and SMEs, Brussels, Belgium
Mohamed.BRAHMI@ec.europa.eu

**Maria Christina Galassi**
European Commission, Directorate General for Internal Market, Industry, Entrepreneurship and SMEs, Brussels, Belgium
Maria-Cristina.GALASSI@ec.europa.eu

**Daniel B Work**
Institute for Software Integrated Systems, Vanderbilt University, Nashville, TN, United States
dan.work@vanderbilt.edu

**Biagio Ciuffo**
European Commission, Joint Research Centre, Ispra, VA, Italy
Biagio.CIUFFO@ec.europa.eu

**ABSTRACT**

This study investigates the assisted lane change functionality of five different vehicles equipped with advanced driver assistance systems (ADAS). The goal is to examine novel, under-researched features of commercially available ADAS technologies. The experimental campaign, conducted in the I-24 highway near Nashville, TN, US, collected data on the kinematics and safety margins of assisted lane changes in real-world conditions. The results show that the kinematics of assisted lane changes are consistent for each system, with four out of five vehicles using slower speeds and decelerations than human drivers. However, one system consistently performed more assertive lane changes, completing the maneuver in around 5 seconds. Regarding safety margins, only three vehicles are investigated. Those operated in the US are not restricted by relevant UN regulations, and their designs were found not to adhere to these regulatory requirements. A simulation method used to classify the challenge level for the vehicle receiving the lane change, showing that these systems can force trailing vehicles to decelerate to keep a safe gap. One assisted system was found to have performed a maneuver that posed a hard challenge level for the other vehicle, raising concerns about the safety of these systems in real-world operation. All three vehicles were found to carry out lane changes that induced decelerations to the vehicle in the target lane. Those decelerations could affect traffic flow, inducing traffic shockwaves.

**INTRODUCTION**

On the road to automation, countries included in the 1958 Agreement are reaching an important milestone with the recent UN Regulation 171 on Driver Controller Assistance Systems (DCAS) (*1*). For the contracting parties, including the European Union, Japan and others, systems of SAE automation level 2 (*2*), have been already available, but their capabilities have been limited to those covered by UN Regulation 79 (*3*). The new regulation treats the system as whole, not separating longitudinal, lateral, and active safety control, and allows for complicated maneuvers, further reducing driver workload. The impact of the system is heightened, both regarding safety and traffic efficiency, as more processes are being automated, still under the control and responsibility of a human driver. The third phase of this regulation is in the drafting stage, relevant to highly complex systems that are not yet available in the market, thus data on good practices and caveats are not existing yet. Due to the different regulation frameworks, systems that closely resemble the most complex versions of DCAS are available in the US market that might not be available in the other regions.

In this light, the European Commission's Joint Research Centre, in collaboration with Tennessee Department of Transportation and the Institute for Software Integrated Systems of Vanderbilt University, have carried out an experimental campaign in the US, to collect data relevant to advanced level 2 systems in the US market, in real world conditions. The assisted lane changing performance of those systems is one of the features that are under investigation. Additional to the regulation efforts, the data collected, and the conclusions drawn, are significant for research on several dimensions relevant to vehicle automation. Improperly executed lane changes can bring a notable increase the risk of collisions (*4*). Furthermore, lane changes can impact traffic flow, initiating traffic shockwaves that are then amplified upstream (*5*), or acting as moving bottlenecks (*6*). As assisted lane changes become more accessible and widely used, their operational characteristics may significantly influence traffic dynamics and stability. Although considerable research and data exist regarding the longitudinal control of current driver assistance systems (*7*, *8*), empirical studies focusing on driver interaction with assisted lane changes are still limited.

The current study presents the first results on investigating the assisted lane change function of five different vehicles, from different manufactures. First, the lane change trajectory is being studied for the actual dynamics, assertiveness, and reproducibility. Then, for three of the vehicles enough data is collected and processed, investigating the safety margins to surrounding traffic. The key question of whether these systems perform lane changes safely is not straightforward to answer. While avoiding defining what is or should be considered as a safe lane change, the assisted lane changes are investigated, comparing them to UN Regulation 171 requirements. Moreover, the Fuzzy Safety Model is used (*9*), to evaluate the challenge level of the cut-in for the vehicle in the target lane. Finally, the deceleration imposed by the safety driver in the other vehicle is obtained from the real-world data, and to evaluate the safety level of the assisted lane changes. One of the systems was shown to perform lane changes with very small gaps, imposing hard decelerations on other vehicles, indicating a potential over-reliance on the system's operation, who ideally should reject such risky maneuvers.

**LITERATURE REVIEW**

While significant progress has been made in understanding the longitudinal behavior of automated vehicles, there remains a notable gap in the empirical study of lateral control, especially lane-changing behavior. Several large-scale datasets have been instrumental in advancing our understanding of longitudinal dynamics. The OpenACC dataset, developed by the Joint Research Centre of the European Commission, provides a comprehensive collection of car-following experiments involving both human-driven and ACC-equipped vehicles across various environments, including public roads, test tracks, and controlled urban networks (*7*). Similarly, the Vanderbilt dataset includes controlled experiments with ACC vehicles responding to speed perturbations, offering valuable insights into string stability and the propagation of traffic waves (*10*, *11*). The experiments referred to as MA and GA, further contribute by analyzing the behavior of multiple commercial ACC systems under disturbance scenarios and near-stop conditions, helping to estimate jam densities and systems' reaction times (*12*, *13*). Additionally, the CARMA platform (*14*), developed by the U.S. Department of Transportation, integrates ACC with vehicle-to-vehicle (V2V) and vehicle-to-infrastructure (V2I) communication to study cooperative driving behaviors in platooning scenarios. Recently, the Third Generation Simulation Data (TGSIM) has been made available, showing that human drivers may change their behavior when around automated vehicles (*15*). These datasets have been critical in validating theoretical models of car-following behavior, assessing the stability of automated traffic streams, and informing policy on the deployment of driver assistance systems. However, despite these advances, similar large-scale, open-access datasets focusing on lateral maneuvers—such as lane-changing—are scarce. Given the importance of lane-changing in shaping traffic flow efficiency and safety, particularly in mixed traffic environments, this lack of empirical data represents a clear gap in the literature and an opportunity for future research.

Few studies have utilized real-world traffic data to investigate the impact of lane-changing behavior on traffic flow dynamics. Mauch and Cassidy (*16*) analyzed a 10-kilometer highway segment with on-ramps and off-ramps, revealing a strong correlation between lane changes near interchanges and the emergence of traffic oscillations. Their findings suggest that, in moderately dense traffic, such oscillations are more likely triggered by random lane changes than by car-following behavior. Cassidy and Rudjanakanoknad (*17*) further explored this phenomenon by examining a freeway segment with an active bottleneck. They identified that capacity drops were primarily caused by vehicles changing lanes from the shoulder to the median lane near merges. Laval and Daganzo (*6*) focused on freeway sections without ramps, where lane changes are typically discretionary. They demonstrated that lane-changing vehicles act as moving bottlenecks while accelerating to match the speed of the target lane, contributing to flow reductions. Ahn and Cassidy (*5*) also provided empirical evidence showing that lane-changing activity significantly influences the formation and propagation of traffic oscillations. These studies underscore the critical role of lane-changing behavior in shaping traffic dynamics and highlight the importance of incorporating real-world observations into traffic flow modeling.

Although lane-changing behavior is known to affect traffic flow and safety, the behavior of assisted and automated systems already available on the market remains largely understudied. This gap is especially critical given the regulatory contrast between Europe, where such systems face strict constraints, and the U.S., where no such limitations exist. Understanding how these deployed systems operate in real-world conditions is essential for evaluating their true impact.

## METHODS
### Experimental campaign
In the fall of 2024, an experimental campaign has been carried out in the US by the Joint Research Centre and the Institute for Software Integrated Systems of Vanderbilt University. For a week, experiments were performed in the I-24 MOTION testbed, a four mile section of Interstate 24 in the Nashville-Davidson County Metropolitan area with 276 ultra-high definition cameras to create a digital twin of the traffic flow (*18*). Each day of the week, six instrumented vehicles run several times in the Interstate 24 section, performing different types of tests.

The vehicles were equipped with the latest versions of level 2 systems, with enhanced features (informally identified as Level 2+). All six vehicles could perform assisted lane changes, in which the system takes over the lateral control to perform the lane change, while the driver responsible for monitoring the system and the environment. Unfortunately, for one of the vehicles, this assisted lane change feature was geofenced, and the specific Interstate 24 section was outside the operational design domain (ODD). All the rest did perform requested, assisted lane changes, in which the driver requested the lane change, and the system performed it. One of the vehicles would also suggest lane changes and perform them after confirmation from the driver. Another one could do what is defined in international regulation as system-initiated, in which the system performs a lane change without the driver requesting or even having to confirm the maneuver.

In each vehicle there was a safety driver and at least one passenger, responsible for the data collection, taking relevant notes, and communicating with the rest of the experiment group. This communication has been invaluable, not just for coordinating the experiment, but also to inform for safety relevant occurrences, as the tests were executed on a freeway in normal operation, around real traffic.

Moreover, a GNSS with IMU solution was installed in each car, to get high accuracy trajectories, and three cameras pointing forward, backward, and to the instrument cluster. Short video recordings are saved when the passenger responsible for annotations, declares an event. Those parts last two-minutes, one before and one after the annotation, and can be used to manually check some events during the post processing, e.g. to identify if a lane change was controlled by the driver or by the system.

One of the experiments was intended to investigate the safety margins for assisted lane changes. One vehicle was performing assisted lane changes, while another of the instrumented vehicles creates a restriction, traveling on the target lane, close to the lane change area. Only four of the vehicles have been involved in those parts. The other two, which were able to perform driver confirmed or system-initiated lane changes, were performing different experiments at the time. For three of the vehicles involved in those tests, considering that one could not perform assisted lane changes on the location, several lane changes have been observed with another vehicle close in the target lane.

## Data processing

Data from the whole week have been collected, synchronized between the different data sources, and interpolated to 10 Hz. Moreover, the data have been organized on laps, with each lap being a trip from the one end to the other of the I-24 MOTION, either eastbound or westbound, gathering also the corresponding annotations from the passengers, and the relevant videos. Moreover, the latitude and longitude of the trajectories are transformed into the I-24 coordinates system, which is used for the openly accessible datasets coming out of the I-24 MOTION testbed (*18*). Using information about the lane markings, the lane of travel is identified for each vehicle and each time stamp. Up to this step, there is no filtering applied in the data, apart from the linear interpolation, to synchronize the different sources.

Having the lane information makes it possible to identify all lane changes and not just the ones in the specific experiments described in the previous subsections. Indeed, manual and assisted lane changes were used by the safety drivers to set the platoon formation for other tests, catch up with the platoon, overtake vehicles when necessary, etc. From all lane changes identified, the data recordings 8 seconds before and 12 seconds after the estimated lane change instant are isolated. Lane changes with an overall lateral displacement less than 3 meters are excluded, as they were mostly coming from system noise, or they could be incomplete maneuvers. Then, a Savitzky–Golay filter is used on the lateral speed, to smooth high frequency noise. On the filtered signal, there is an additional check for the lateral speed starting from a zero value, or negative, in the direction of the lane change, and ending to a zero value, or negative. This check discards many double lane changes, or even manual lane changes that are carried out in steps, with the driver "negotiating" for the space in the target lane.

After the checks, 692 lane changes, for all six vehicles are identified for which, using the notes and videos recorded, the lane changes are divided into manual, assisted with the driver requesting, assisted with the driver confirming, system initiated. For one of the cars, we identified a few interrupted cases. Those were assisted, requested by the driver, but the driver took back control of the steering wheel before the maneuver was all the way through. Looking into the videos, we do not see any safety reason relevant for the interruption, so it was possibly the driver being impatient. The overall number of cases is presented in **Figure 1**. Cars D and E can perform user confirmed and system-initiated maneuvers respectively. Moreover, those two have not been involved in the tests of the lane change safety margins, but performed more lane changes, in front of real traffic, during the same time. Finally, car F did not perform assisted lane changes in I24 as it was outside of the specific feature's ODD.

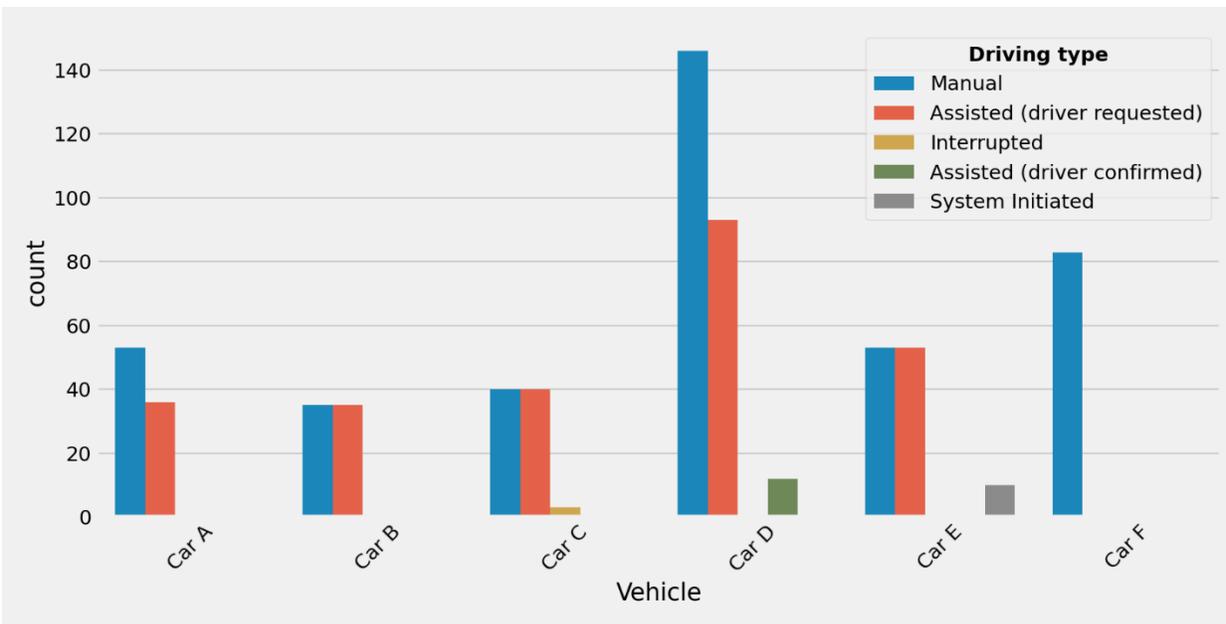

**Figure 1 Number of lane changes per vehicle and driving type**

For the investigation of the safety margins, all cases of vehicles lane changing in front of another instrumented vehicle being closer than 100 meters on the target lane are isolated, regardless of the level of automation or the type of experiment during which they occurred. Therefore, for the safety relevant investigation the number of lane changes dropped to 161. The number of cases per automation system and vehicle are presented in **Figure 2**. The labels and colours are the same as **Figure 1**. For cars D and E, not involved in the relevant tests, there is not enough data for the safety relevant investigation. However, the I24 camera data can be used in future work, to include cases of lane change in front of real traffic.

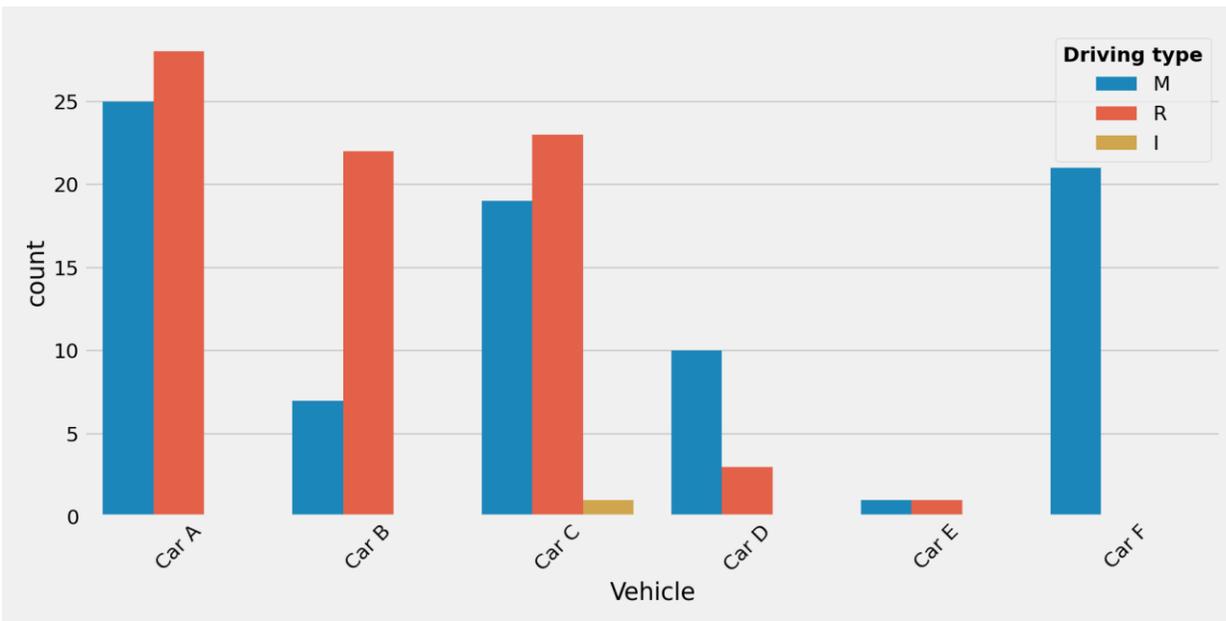

**Figure 2 Number of lane changes per vehicle and driving type with a following vehicle. "M" denotes manual lane changes, "R" denotes assisted (driver requested) lane changes, and "I" denotes interrupted lane changes.**

**Safety relevant evaluation**
The safety level of the lane changes may not always be straightforward to evaluate. In this work, we investigate the lane changes in three different approaches. First, we compare the distance at the beginning of the lane change, comparing to the critical distance in relevant UN regulations such as UN Regulation 79 (*3*), which has been used for the approval of the vast majority of ADAS systems with assisted lane change in the European market. The critical distance defined in UN Regulation is echoed in the more recent UN Regulation 171, concerning level 2 systems with complex capabilities, defined as Driver Control Assistance Systems (DCAS) (*1*), that better resembles the systems in the vehicles used. For DCAS the critical distance is defined as:

$$s_{171} = (u_{rear} - u_{DCAS})t_{reaction} + \frac{(u_{rear} - u_{DCAS})^2}{2a_{rear}} + u_{DCAS} * t_G, \qquad (1)$$

where $u_{rear}$ and $u_{DCAS}$ the speeds of the rear vehicle and the DCAS vehicle, $t_{reaction}$ the time for the other vehicle to react, set to be 1.4 s , $a_{rear}$ the deceleration of the rear vehicle is set to 3 m/s², and $t_G$ the safe time gap of 1 s. The distance is calculated at the time that the DCAS system starts the lane changing manoeuvre. It has to be noted that the vehicles tested are not subject to any of those regulations, as the legal framework is different in the US. However, especially from a European perspective, it is imperative to understand the behaviour of such systems where they are not strictly regulated, as this can produce valuable feedback.

Additionally, a simulation model is used to classify the safety level of the lane change, that reproduced the reaction to an attentive human driver to a cut-in case. The Fuzzy Safety Model (FSM) model was initially developed to classify preventable and unpreventable cut-ins, for the vehicle receiving the cut in (*9*). However, it is also able to classify the challenge level of a manoeuvre. Thus, the FSM is currently used, to evaluate the assisted lane change safety level, by the classification of the challenge level to the vehicle in the target lane. Finally, the immediate response of the following vehicle is examined, looking at the maximum deceleration that the safety driver applied. Drivers are known to relax their desired safety margins temporarily after lane changes (*19*), so the real deceleration triggered might differ from the simulated. Those real decelerations due to improper, in some cases, lane change, can potentially induce traffic shockwaves.

**RESULTS**
The results regarding the characterization of assisted lane changing are divided into two parts, the lane changing dynamics and the safety margins.

**Lany changing dynamics**
In the lane changing dynamics, the lateral speeds, acceleration, and consistency between different lane change trajectories are investigated. In **Figure 3a** the boxplots of the lane change duration for the different vehicles and different types of assistance. The blue boxes, representing manual lane changes, are shown to be on average longer in duration, than the red ones, representing the assisted lane changes requested by the driver. Interrupted lane changes have a longer duration, possibly because of the shift in control. Car D, that can suggest lane changes, and get the driver's confirmation, is managing the quickest lane changes, dropping below 6 seconds for the whole maneuver. Moreover, the duration seems to be even more consistent than the rest of the systems. The system initiated by Car E, seems to produce a heavier tail towards longer durations than the requested lane changes for the same car. This can be expected, as for those cases, the time for the driver to realize the lane change is happening should also be considered by the developer of the system.

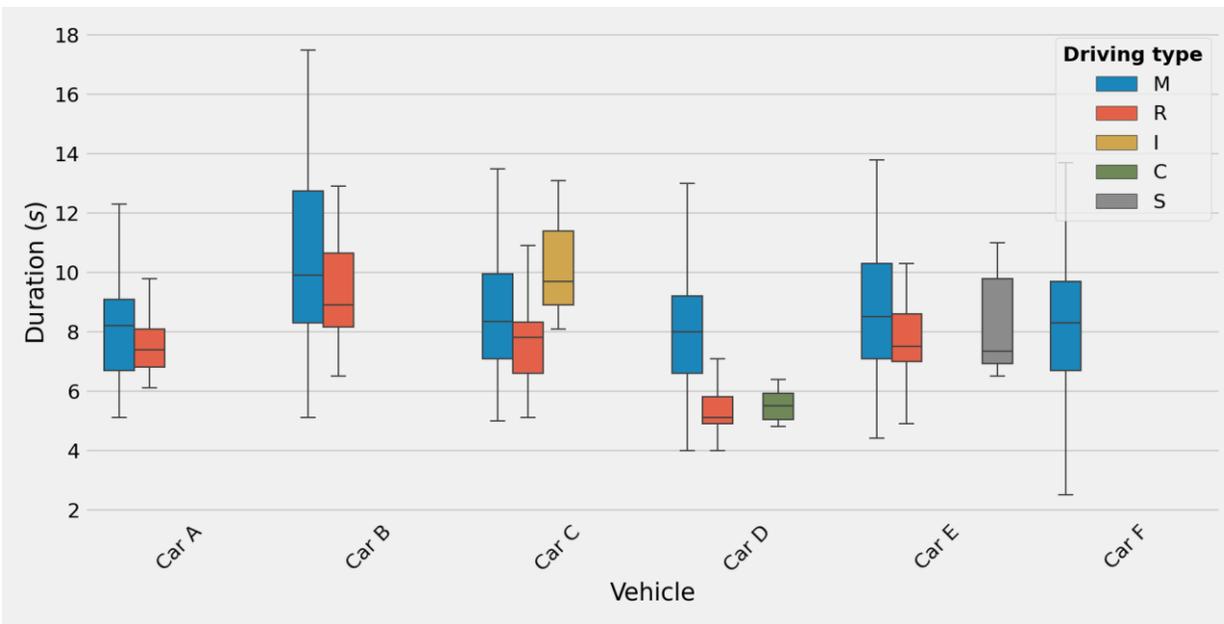

(a)

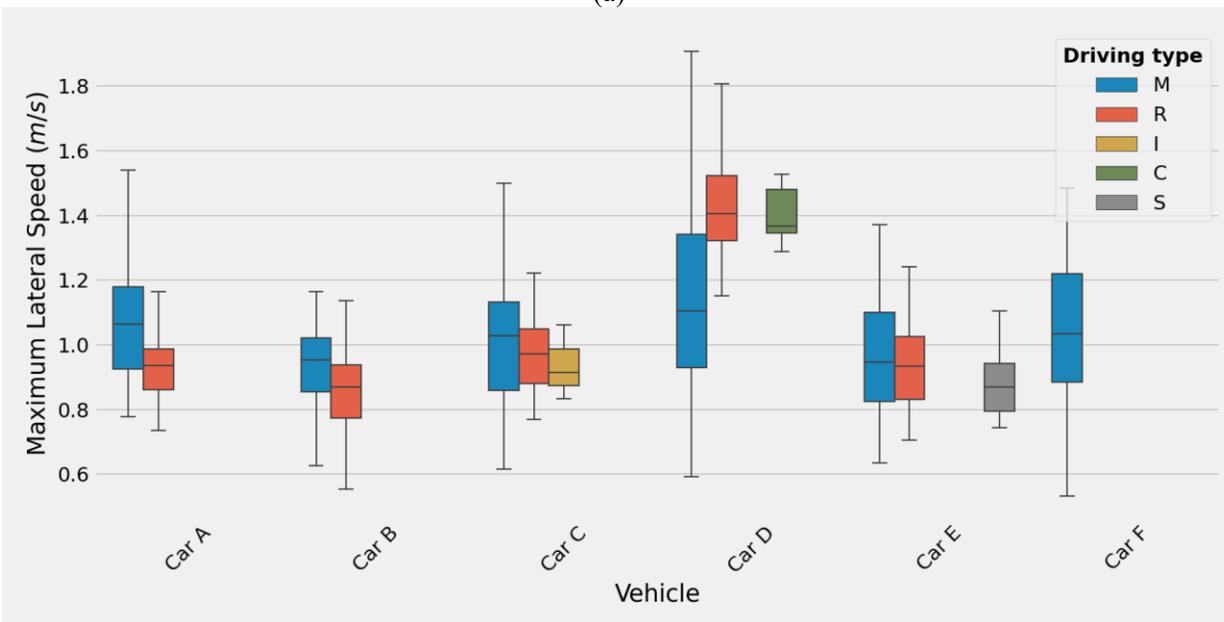

(b)

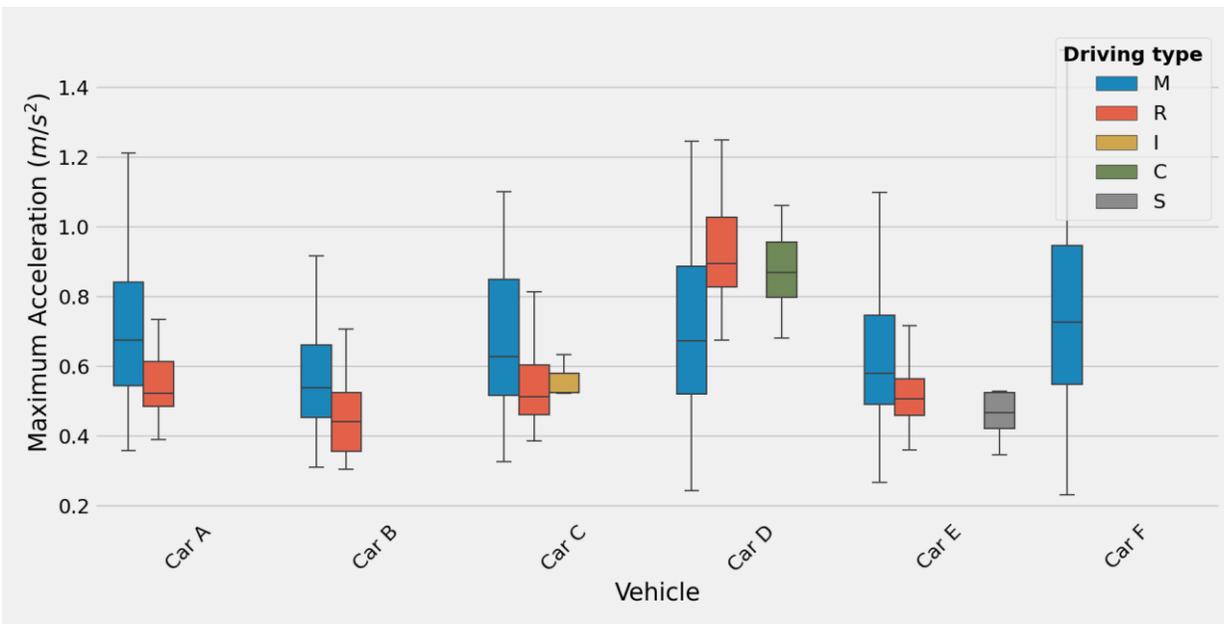

(c)

**Figure 3 Lane change a) duration, b) maximum lateral speed, c) maximum acceleration. "M" denotes manual lane changes, "R" denotes assisted (driver requested) lane changes, "I" denotes interrupted lane changes, "C" denotes assisted (driver confirmed) lane changes, and "S" denotes system initiated lane changes.**

In **Figure 3b** and **Figure 3c** the maximum lateral speed and maximum acceleration are presented. Most ADAS systems seem reluctant to use similar maximum lateral speed and acceleration as the human driver. There is an obvious exception with car D using speeds and accelerations faster than most of the safety drivers. The system's developer appears to prioritize executing a lane change as soon as it is deemed safe, doing so quickly and assertively to preempt any changes in the surrounding environment. Interestingly, the safety drivers, which have been rotating between cars, or the passengers, never expressed concerns or discomfort about this fast lane change being uncomfortable or unsafe.

Regarding the 5 vehicles that were able to perform assisted lane change, their trajectories are compared, by investigating the lateral speeds used, as presented in **Figure 4**. The lateral speed evolution in time is very consistent for the assisted lane changes, regardless of the longitudinal speed, that varies. Moreover, it shows to be consistent although driver requested, driver confirmed, and system initiated lane changes are grouped. Apart from a few outliers that may have been misclassified in the manual process, the rest of the trajectories are following a sinusoidal function, which cannot be said for manual lane changes. Furthermore, the higher lateral speeds and accelerations employed by car D are confirmed.

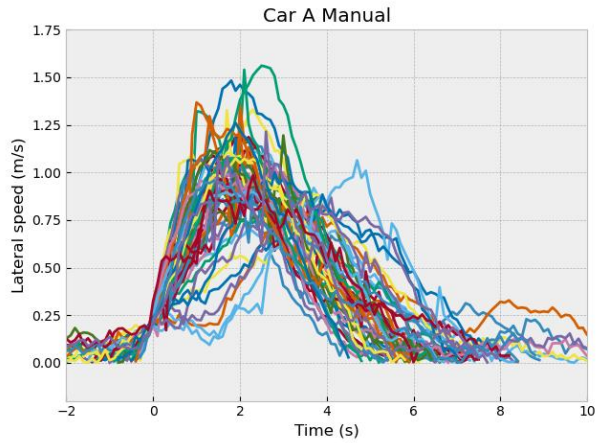
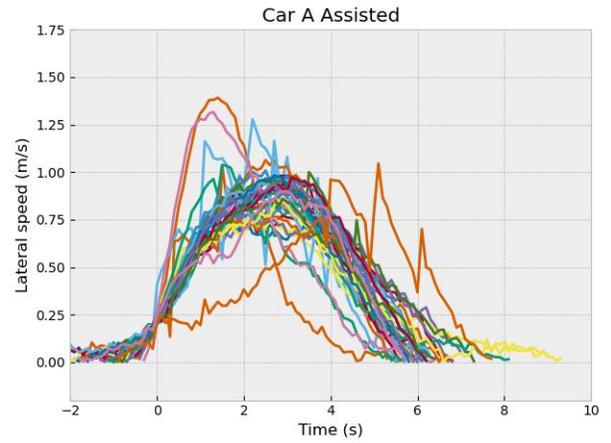
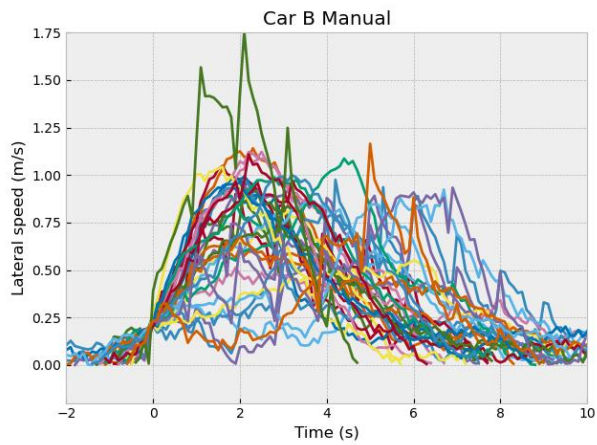
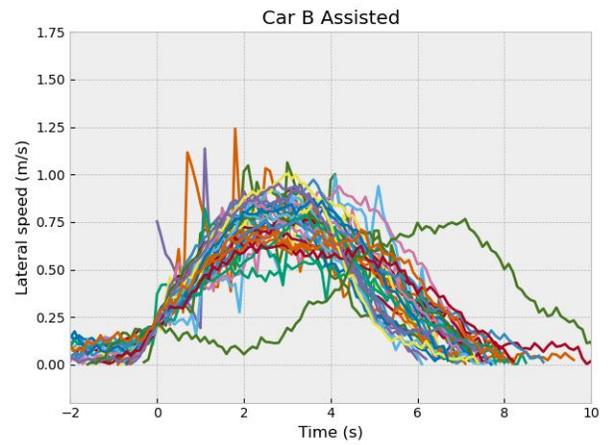
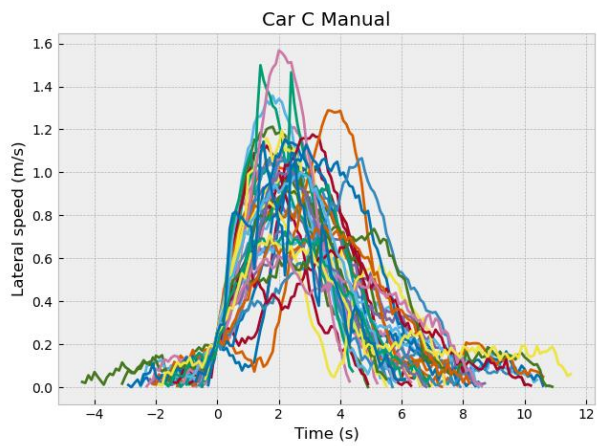
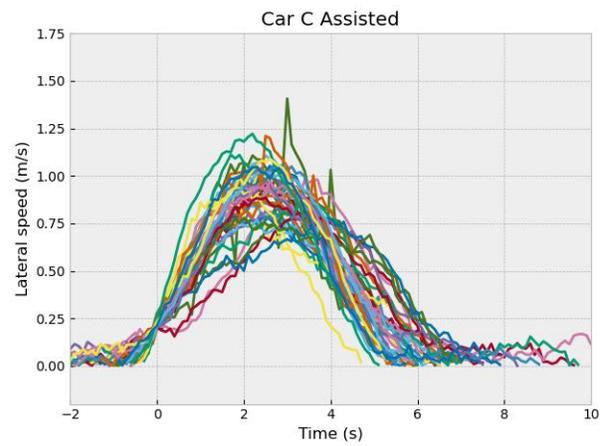

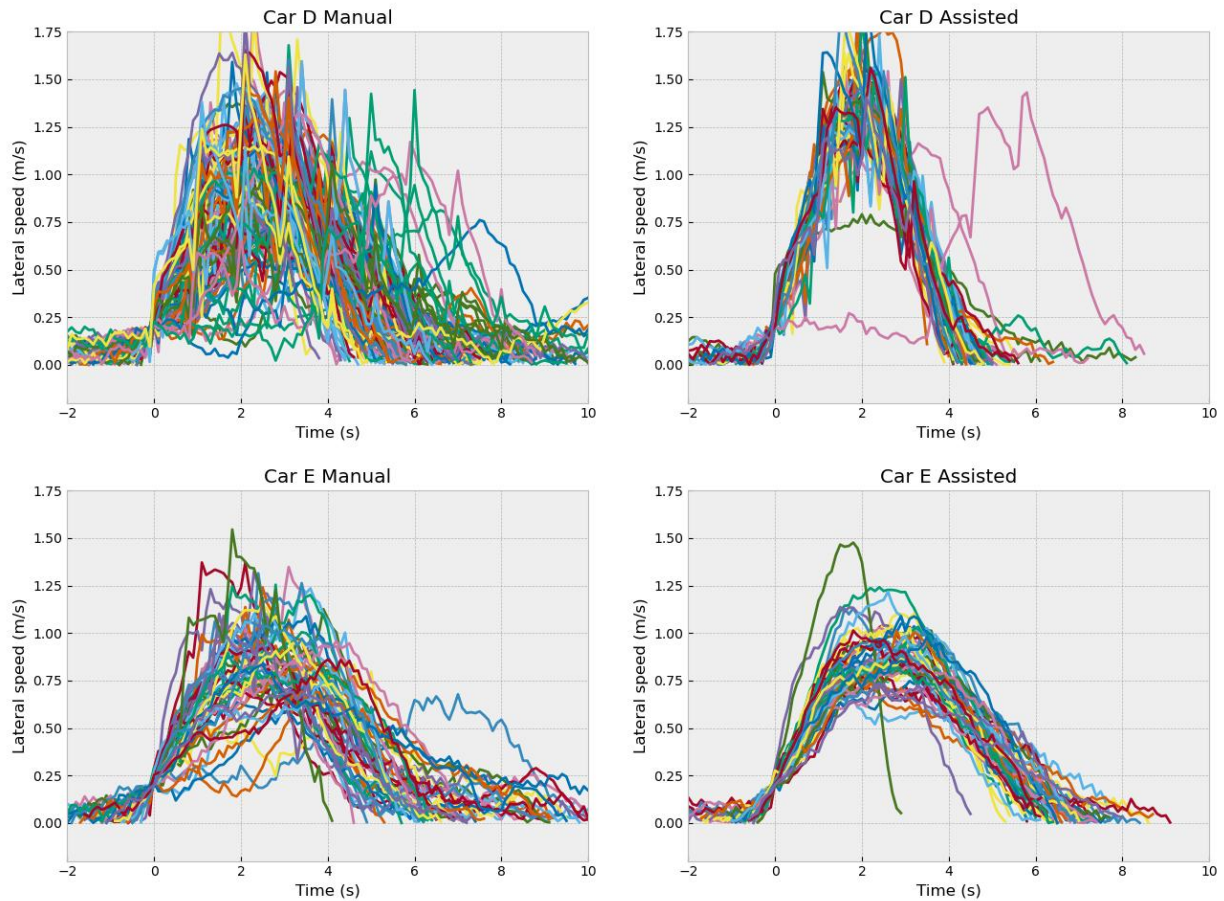

**Figure 4 Lateral speed trajectory for the different vehicles and driving types**

To evaluate the difference in the lane change speed consistency, and error metric has calculated. For each of the 10 different cases presented in **Figure 4**, the median value is found for each 0.1 second. Then the absolute error of each individual lane change is integrated, along the lane change duration. The resulting histograms are presented in **Figure 5**, for manual and assisted cases. For the assisted case, the majority of lane changes have an integrated error that is less than 1 meter, while for manual lane changes the peak of the histogram is around 1 meter.

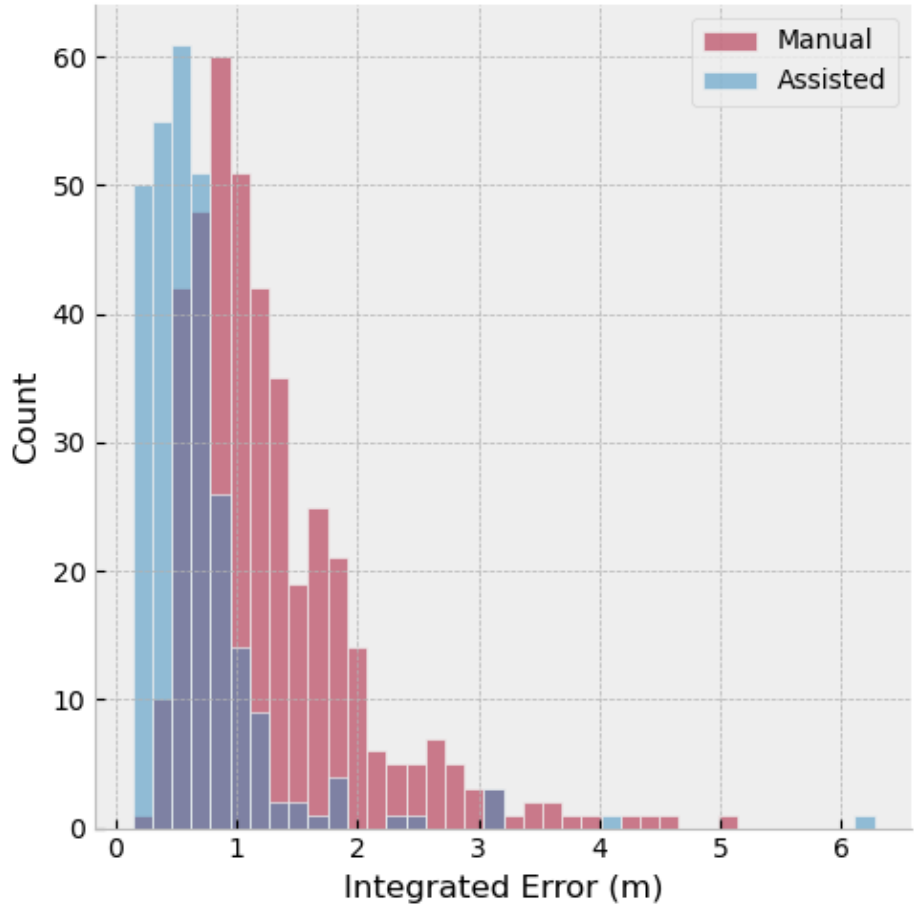

**Figure 5 Integrated error compared to the median lane change for manual and assisted lane changes**

Considering the existence of outliers in **Figure 5**, and some trajectories observed, that seem misclassified, a simple filtering algorithm is performed, discarding assisted lane changes with an error value that is larger than 1.5 meters. The results for two vehicles are presented in **Figure 6a** and **b**. The blue lines represent the accepted trajectories, the black line is the median value, and the red lines represent the trajectories filtered out of the assisted lane changing class. Thus, the consistency of the assisted lane changes can help identify lane changes that are not assisted, even with a simple algorithm.

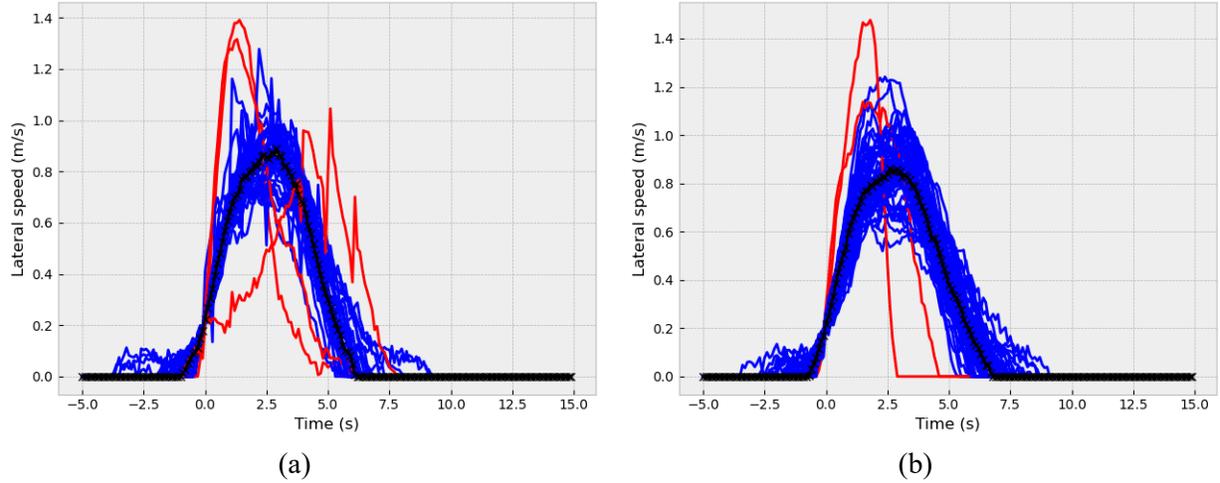

(a)                                                   (b)

**Figure 6 Filtering results for Car A, and Car E**

Considering the reproducibility of the lane changing behavior, we attempted to use a sinusoidal function to replicate it. After iteration and different tries, the need to use an asymmetric function was clear, in order to consider a small asymmetry between the accelerating and decelerating branches. Thus, we define a parametric cosine-shaped function that is asymmetric about a central peak. The function is constructed using two half-cosine lobes with independently tunable widths on the left and right sides of a central point.

$$f(x) = \begin{cases} \dfrac{a}{2}\left[1 + \cos\left(\pi\dfrac{x-c}{w_\ell}\right)\right], & \text{if } c - w_\ell \leq x < c \\ \dfrac{a}{2}\left[1 + \cos\left(\pi\dfrac{x-c}{w_r}\right)\right], & \text{if } c \leq x < c + w_r \\ 0, & \text{otherwise} \end{cases} \tag{2}$$

where $a$ is the amplitude, $c$ is the center, $w_\ell$ the width of the left half, and $w_r$ the width of the right half.

The fitted lines are able to capture the asymmetry in the data, that Gaussian or standard cosine functions would not be able to. The results of fitting this curve are presented in **Figure 7**, with the red lines, while the original data are denoted with blue lines, and the median value speeds with green dots. The fitted function does capture the dynamics, with the modelling error appearing smaller than the noise in the data. The fitted values for each vehicle are presented in **Table 1.** The difference between $w_l$ and $w_r$ is a quantification of the asymmetry between the two branches of the lateral maneuver. Smaller $w_l$ values show a slight increase in the urgency to start the movement and claim the space, and the larger $w_r$ values show a calmer approach in finishing the movement and placing the vehicle correctly in the center of the target lane. Again, car D shows to be the exception, with this asymmetry being subtle.

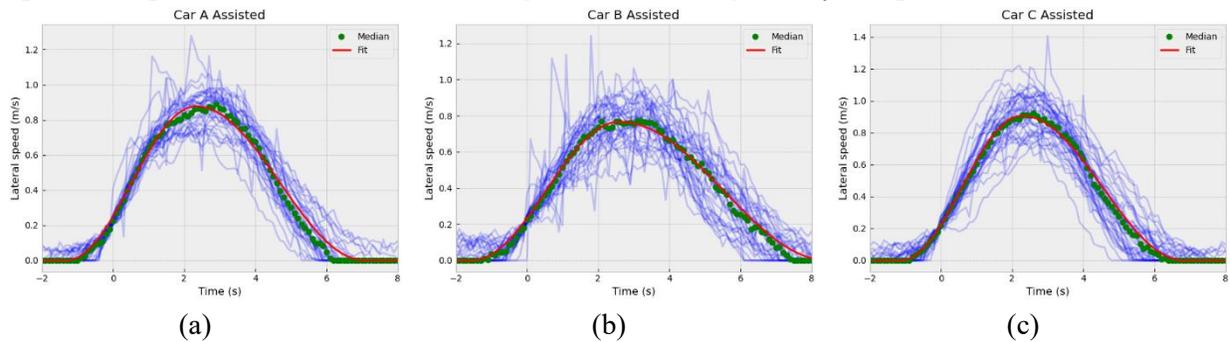

(a)                                   (b)                                   (c)

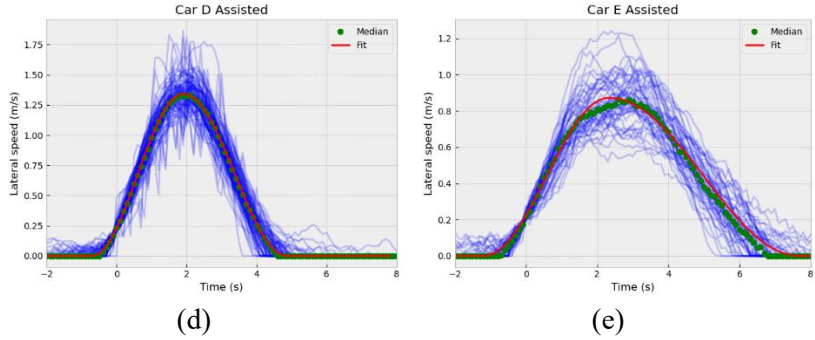

**Figure 6 Fitted curves of the asymmetric cosine**

**TABLE 1 Asymmetric cosine values fitted on the lane changes**

| **Vehicle** | $a$ | $c$ | $w_l$ | $w_r$ |
|---|---|---|---|---|
| **Car A** | *0.877* | *2.336* | *3.644* | *4.745* |
| **Car B** | *0.766* | *2.605* | *4.151* | *5.872* |
| **Car C** | *0.905* | *2.286* | *3.415* | *4.557* |
| **Car D** | *1.340* | *1.905* | *2.619* | *2.970* |
| **Car E** | *0.871* | *2.332* | *3.532* | *5.354* |

**Safety margins**

The investigation on the safety margins of the lane change is relevant only for cars A, B, and C, as only for those the collected data is sufficient. The first step in the safety evaluation is to inspect the distance to the vehicle in the target lane, at the instance that the vehicle under test starts the lateral movement and compare to the critical distance according to the UN Regulation 171. The results of this check are presented in **Figure 7**. The red parts of the bars represent the number of times the lane change started from a distance that is not safe according to the regulatory requirement. The hatched bars present the results for assisted driving, while the plan bars report the results for manual lane changes. The results show that both human drivers, and assistance systems investigated, are carrying out lane changes with distances smaller than the requirement. Of course, those systems are not regulated under 171. Moreover, Car C seems to be the most conservative, with a larger ratio of compliant maneuvers, compared to the other two.

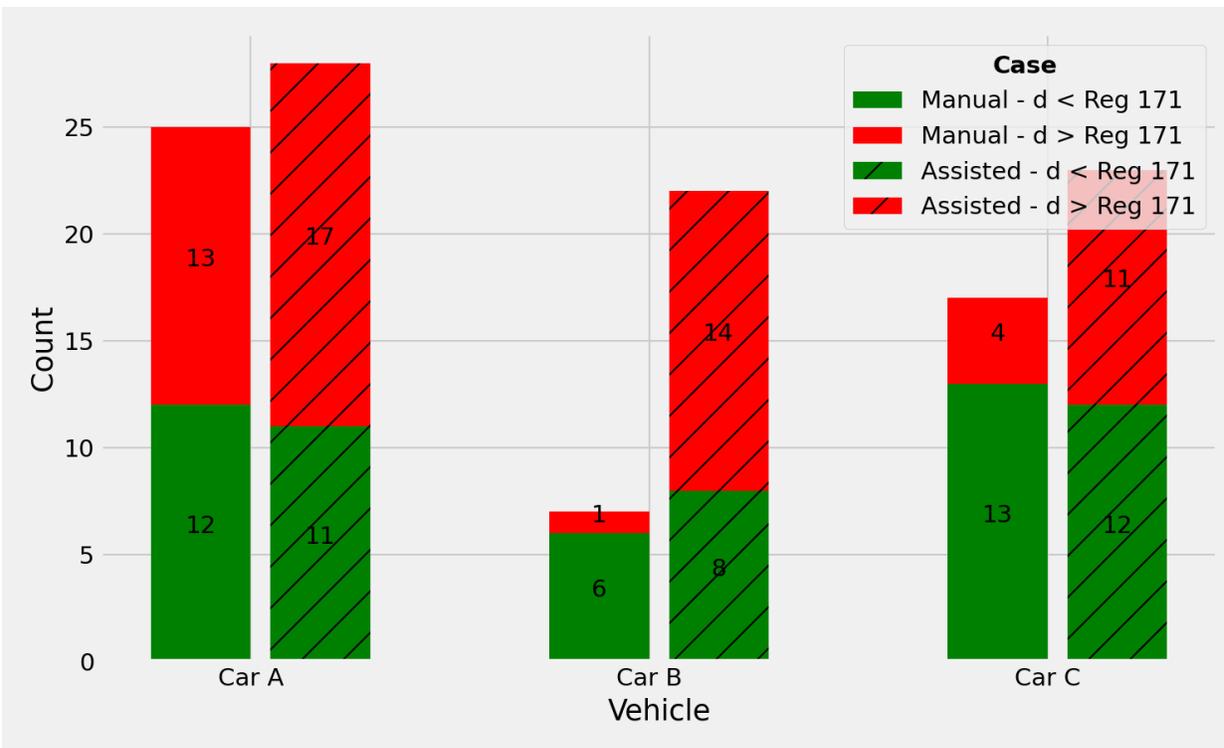

**Figure 7 Comparison to UN Reg 171 critical distance**

To further investigate the safety level of the lane changes recorded, the FSM model was used (*9*). The simulation reproduces the reaction of an attentive driver, and the cases are classified according to the challenge level. The results are presented in **Figure 8**. Easy, medium and hard challenge level are represented by green, yellow, and red color, respectively. Medium challenge level cases are cases in which the driver of the vehicle receiving the lane change must decelerate moderately to keep the safety distance. For the hardest cases, strong deceleration is required to avoid an accident. Again, both human drivers, and assistance systems, are not avoiding forcing deceleration to vehicles upstream, with car C being the least aggressive by a small margin. What is important is the hard case observed for car A for an assisted lane change. Such strong cut-ins are not common even in real traffic and trigger a strong response from the other vehicle. The subjective feeling of the team of researchers during the experiment is confirmed by the results, as occupants of the relevant vehicles clearly indicated that the situation at the time was unsafe. Such a harsh lane change was not observed by any other of the systems. Moreover, for car A, we did not try to replicate the situation, out of concern for the actual safety of the experiment, that was taking place on real roads. However, since the system is quite deterministic in its behavior, it can be assumed that such lane changes happen in the real world. It is the responsibility of the driver to monitor, realize the risk, and take over, but driver overreliance could decrease the efficacy of the driver as a safety guarantee.

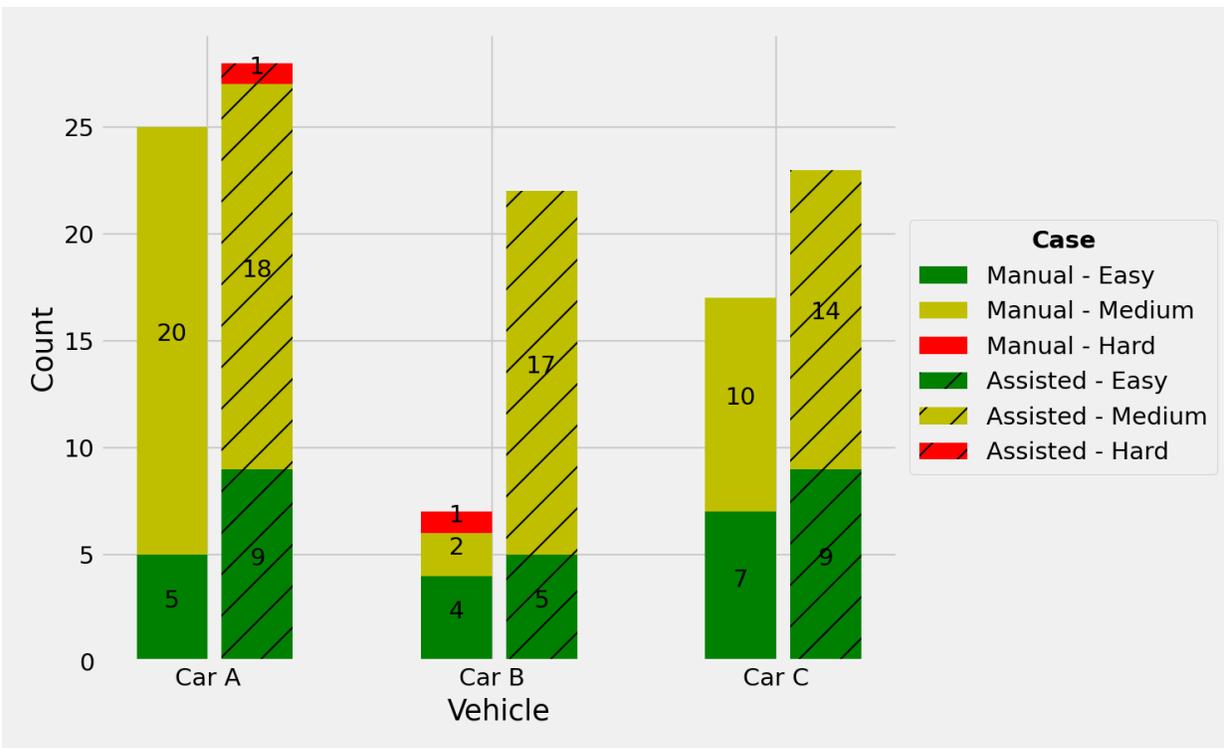

**Figure 8 FSM results for the observed lane changes**

Finally, the actual reaction of the safety driver receiving the lane change is inspected. The results are presented in **Figure 9** for the assisted lane changes. Green bars denote when the human driver is not decelerating harder than 0.5 m/s², so no substantial deceleration. If the maximum deceleration recorded is harder than 3 m/s², this is denoted as a hard deceleration reaction, as those deceleration magnitudes are uncommon in highways. For cases in between, the deceleration is described as calm, represented by the yellow bars. Firstly, the specific event of car A is still noticeable as the only hard deceleration case. Moreover, by the rate of no deceleration cases, it is shown that the other two systems are more conservative than that of car A. Moreover, the green parts are larger here, compared to the regulatory critical distance and the FSM results, indicating the relaxation behavior of human drivers, that temporarily accept short gaps, to avoid a strong and abrupt reaction.

Apart from the direct safety impacts, it is worth noting that even the calm decelerations due to short lane changes are perturbations on the traffic flow (*5*). In congested or close to congested conditions, those perturbation travel upstream, with their magnitude often increasing due to string instability (*20*). Even for car C, which is shown to be the more restricted of the three, there are often cases of the system lane changing in a way that forces a reaction from the new follower vehicle. Further research is necessary to investigate the exact mechanisms and identify optimal for safety and efficiency lane changing operational limits.

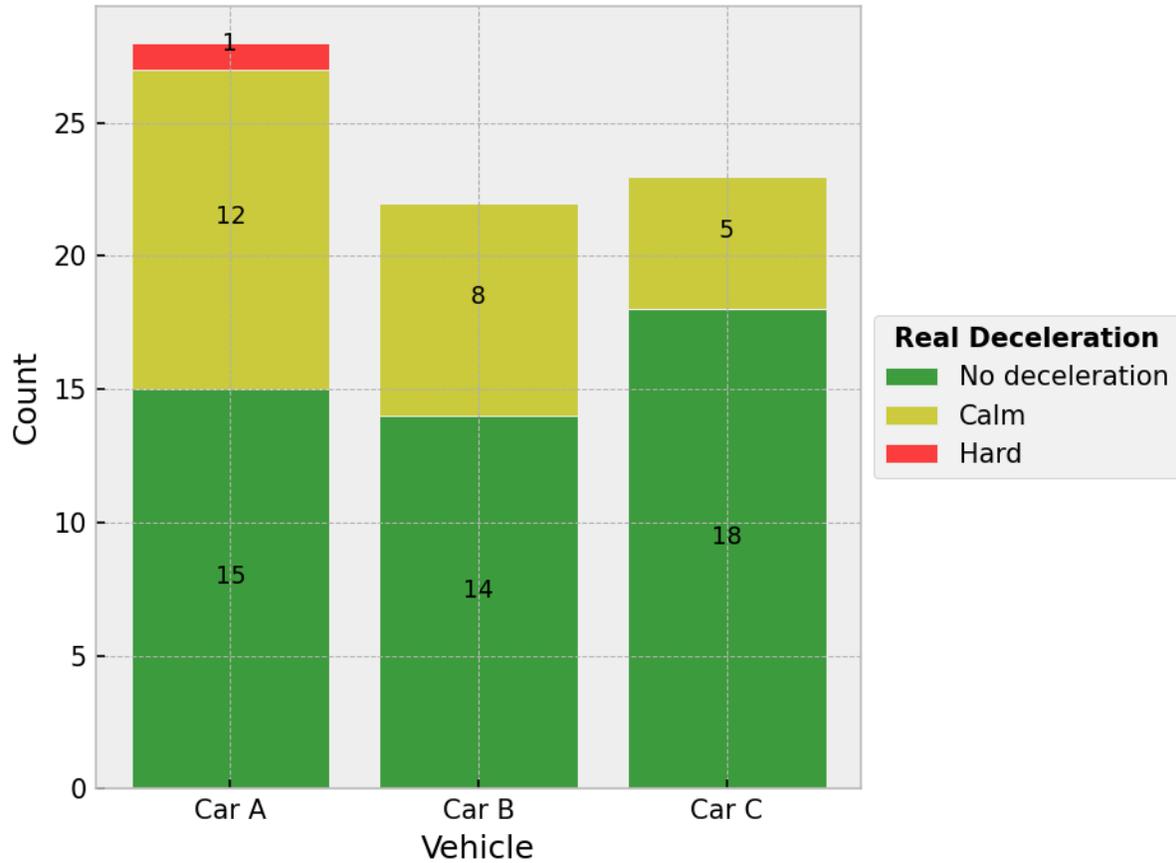

**Figure 8 Real driver reactions**

**CONCLUSIONS**

Driving automation systems with low levels of automation are becoming increasingly available, and more complex in their capabilities, compared to a few years ago. Ongoing research aims to predict and evaluate potential opportunities, limitations, and the broader effects of these systems on traffic networks. Advanced Driving Assistance Systems (ADAS) that provide longitudinal control have been widely studied for their impacts. Empirical data from ADAS systems available on the market have been significant in clarifying the next challenges and opportunities. However, there is a lack of understanding of the characteristics of ADAS systems that provide lateral control, as those systems are less widely available, and in some cases, more strictly regulated.

Lane changes on highways are critical to both road safety and the dynamics of traffic flow. When executed improperly, they can introduce disturbances or function as moving bottlenecks.(*6*). The current work presents the results of an experimental campaign that are relevant to ADAS systems performing lane changes. Five different cars capable of assisted lane changes, are used, and this feature is investigated for the relevant kinematics. Furthermore, for three of the vehicles, the safety level of the lane changes performed is investigated.

Results show that the kinematics of assisted lane changes are very consistent for each system, and as such they are easy to be identified and replicated. Moreover, regarding the kinematics, four out of five vehicles used slower lateral speeds and decelerations than a human driver, with only one consistently being more assertive in performing a lane change when enough space is identified. This system performed lane changes that had a duration of around 5 seconds, compared to other systems needing 7 seconds or more for the whole maneuver. This shows an important divergence in the planning and execution of assisted lane changes, that may be required to be investigated for the safety and traffic flow impacts.

Regarding the safety margins, the vehicles sold and operated in the US, are not restricted by EU and UNECE regulations, and they are not designed to this regulatory requirement. This is confirmed by comparing the distances to the relevant UN Regulation 171. A simulation method used to classify the challenge level for the vehicle in the adjacent lane, receiving the lane change, showing that those systems might induce decelerations on the other vehicle, in order to keep safe distances. Moreover, one of the assisted systems carried out a maneuver that was found to present a hard challenge level for the new follower vehicle. This observation is concerning, as such systems are currently operated in real public roads by users. Moreover, in the data from the experiments, assisted lane changes commonly trigger decelerations on the surrounding vehicles, such that are known in the literature to be able to initiate traffic shockwaves. Further research is necessary to investigate the exact mechanisms and identify optimal for safety and efficiency lane changing operational limits

**ACKNOWLEDGMENTS**
The study has been carried out in the framework of the Administrative Agreement n° 36948 between Directorate-General Internal Market, Industry, Entrepreneurship and SMEs (DG GROW) and the Joint Research Centre of the European Commission (JRC).

**AUTHOR CONTRIBUTIONS**
The authors confirm contribution to the paper as follows: study conception and design: all authors; data collection: all authors; analysis and interpretation of results: K.M, S.V, G.Z, J.J; draft manuscript preparation: K.M, A.K, M.B, M.G, D.W. All authors reviewed the results and approved the final version of the manuscript.

**DECLARATION OF CONFLICTING INTERESTS**
The authors declared no potential conflicts of interest with respect to the research, authorship, and/or publication of this article.

**FUNDING**
This research has been funded by the Joint Research Centre for the European Commission, Ispra (VA), Italy and by the Directorate General for Internal Market, Industry, Entrepreneurship and SMEs for the European Commission, Brussels, Belgium